\documentclass[aps,prl,showpacs,twocolumn,amsmath,10pt,superscriptaddress,floatfix,nofootinbib,a4paper]{revtex4-1}
\usepackage{epsfig,amssymb,amsfonts,bm,color}
\usepackage{hyperref}

\newcommand{\be}{\begin{equation}}
\newcommand{\ee}{\end{equation}}
\newcommand{\jp}{J/\psi}
\newcommand{\jpsip}{\psi(2S)J/\psi}
\newcommand{\jpsipp}{\psi(3770)J/\psi}
\newcommand{\cccc}{cc\bar c\bar c}
\newcommand{\chisq}{\chi^2/{\rm dof}}

\newcommand{\bonn}{\affiliation{Helmholtz-Institut f\"ur Strahlen- und Kernphysik and Bethe Center for Theoretical Physics,\\ Universit\"at Bonn, D-53115 Bonn, Germany}}

\newcommand{\fzj}{\affiliation{Institute for Advanced Simulation, Institut f\"ur Kernphysik and J\"ulich Center for Hadron Physics, Forschungszentrum J\"ulich, D-52425 J\"ulich, Germany}}

\newcommand{\itp}{\affiliation{CAS Key Laboratory of Theoretical Physics, Institute of Theoretical Physics, Chinese Academy of Sciences,\\ Zhong Guan Cun East Street 55, Beijing 100190, China}}

\newcommand{\ucas}{\affiliation{School of Physical Sciences, University of Chinese Academy of Sciences, Beijing 100049, China}}

\newcommand{\itep}{\affiliation{Institute for Theoretical and Experimental Physics NRC ``Kurchatov Institute'', Moscow 117218, Russia }}

\newcommand{\lebedev}{\affiliation{P.N. Lebedev Physical Institute of the Russian Academy of Sciences, 119991, Leninskiy Prospect 53, Moscow, Russia}}

\newcommand{\mipt}{\affiliation{Moscow Institute of Physics and Technology, 141700, Institutsky lane 9, Dolgoprudny, Moscow Region, Russia}}

\graphicspath{{Figs.dir/}}

\begin{document}

\title{
Coupled-channel interpretation of the LHCb double-$J/\psi$ spectrum and hints of a new state near the $J/\psi J/\psi$ threshold
}

\author{Xiang-Kun Dong}
\itp\ucas

\author{Vadim Baru}
\bonn \itep \lebedev

\author{Feng-Kun Guo}\email{fkguo@itp.ac.cn}
\itp \ucas

\author{Christoph~Hanhart}
\fzj

\author{Alexey Nefediev}
\lebedev
\mipt

\begin{abstract}

Recently, the LHCb Collaboration reported pronounced structures in the invariant mass spectrum of $J/\psi$-pairs produced in proton-proton collisions at the Large Hadron Collider. In this Letter, we argue that the data 
can be very well described within two variants of a coupled-channel approach employing $T$-matrices consistent with unitarity: (i) with just two channels, $\jp\jp$ and $\jpsip$, as long as energy-dependent interactions in these channels are allowed, or (ii) with three channels $\jp\jp$, $\jpsip$ and $\jpsipp$ with just constant contact interactions. 
Both formulations hint at the existence of a near-threshold state in the $\jp\jp$ system with the quantum numbers $J^{PC}=0^{++}$ or $2^{++}$, which we refer to as $X(6200)$. We suggest experimental tests to check the existence of this state and discuss what additional channels need to be studied experimentally to allow for 
distinctive tests between the two mechanisms proposed. 
If the molecular nature of the $X(6200)$, as hinted by 
the three-channel approach,
is confirmed, many other double-quarkonium states should exist driven by the same binding mechanism. In particular, there should be an $\eta_c\eta_c$ molecule with a similar binding energy. 

\end{abstract}

\maketitle

{\it Introduction.}---Quantum chromodynamics (QCD) is highly nonperturbative at low energies. As a result, how hadrons emerge from QCD 
and how the hadron spectrum is organized are still challenging open questions.
The quest of exotic hadrons beyond the conventional quark model classification of quark-antiquark mesons and three-quark baryons has been one of the central issues in the study of nonperturbative QCD. 
In the past decades, dozens of new resonant structures with exotic properties were reported by various experiments 
in particular in the spectrum of hadrons containing at least one heavy-flavor (charm or bottom) quark. However, these 
observations brought even more challenges as they seem not to fit into a single uniform classification scheme, 
and various interpretations were proposed for each of them, see Refs.~\cite{Hosaka:2016pey,Lebed:2016hpi,Esposito:2016noz,Guo:2017jvc,Olsen:2017bmm,Liu:2019zoy,Brambilla:2019esw,Guo:2019twa,Yang:2020atz,Zyla:2020zbs} for recent reviews of such exotic states. 
Recently, the LHCb Collaboration reported resonant structures in the double-$\jp$ invariant mass distribution using data for $pp$ collisions at the c.m. energies 7, 8, and 13~TeV~\cite{Aaij:2020fnh}.
The form of the signal is nontrivial, departing 
significantly from the expected phase space distribution as well as single and double-parton scattering: An enhancement in the near-double-$\jp$ threshold region from 6.2 to 6.8~GeV is seen, which is followed by a narrow peak around 6.9~GeV. Between the broad bump and the narrow peak, there is a dip at  around 6.8~GeV. The narrow peak is now dubbed $X(6900)$, and has spurred a flood of model explanations \cite{liu:2020eha,Wang:2020ols,Jin:2020jfc,Yang:2020rih,Lu:2020cns,Chen:2020xwe,Wang:2020gmd,Sonnenschein:2020nwn,Albuquerque:2020hio,Giron:2020wpx,Maiani:2020pur,Richard:2020hdw,Wang:2020wrp,Chao:2020dml,Maciula:2020wri,Karliner:2020dta,Wang:2020dlo}. Naturally, a fully-charmed compact tetraquark resonance is the most straightforward candidate. However, most of the theoretical studies indicate that the $\cccc$ ground state should have a mass lower than 6.9~GeV~\cite{Iwasaki:1975pv,Chao:1980dv,Badalian:1985es,Ader:1981db,Wu:2016vtq,Karliner:2016zzc,Wang:2017jtz,Liu:2019zuc,Bedolla:2019zwg,Chen:2020lgj}. Furthermore, the 700~MeV energy gap between the double-$\jp$ threshold and 6.9~GeV is larger than a typical energy gap between the ground and radially/orbitally excited states. Thus, lower states should exist, if there is a $\cccc$ resonance with a mass around 6.9~GeV. Due to a smaller phase space, such lighter states are expected to have smaller widths. However, there are no obvious narrower peaks in the reported double-$\jp$ spectrum.

It is well-known that threshold effects play an important, sometimes crucial, role for the properties of hadrons residing above open-flavor thresholds. 
For example, there is always a cusp at an $S$-wave threshold due to the analytic structure of the two-body Green's function (for a review, see Ref.~\cite{Guo:2019twa}). It may lead to either a peak or a dip, depending on the interference with other contributions. The visibility of the corresponding structure in the line shape depends on whether or not it is enhanced by a nearby pole in the amplitude \cite{Guo:2014iya}. Thus, in order to properly interpret the new observations it is important to understand the role played by various thresholds located nearby. There are quite a few double-charmonium channels with the thresholds below 7.2~GeV which can couple to the double-$\jp$ system, such as $\eta_c\eta_c$, $h_ch_c$, $\chi_{cJ}\chi_{cJ'}$ ($J,J'=0,1,2$), {$\eta_c\eta_c(2S)$,} $\jpsip$ and $\jpsipp$.
{In this work we assume that the interaction between the quarkonia is dominated by the exchange of light modes (soft gluons or, e.g., pion pairs).
Then the  coupling of the double-$\jp$ to the $\eta_c\eta_c$ or $h_ch_c$} flips the charm-quark spin, and is expected to be suppressed due to the heavy quark spin symmetry (HQSS).
{Indeed, HQSS implies that the interactions involving the spin of a heavy quark are suppressed as $\mathcal{O}(\Lambda_{\rm QCD}/m_Q)$~\cite{Manohar:2000dt}, where $\Lambda_{\rm QCD}$ denotes the momentum scale where QCD gets nonperturbative and $m_Q$ is the heavy-quark mass.} 
From the point of view of the meson-exchange picture, the lowest meson that can be exchanged for the coupling of the $\chi_{cJ}\chi_{cJ'}$ to the double-$\jp$, keeping the SU(3) flavor and isospin symmetries, is the $\omega$. It is heavier than the $f_0(500)$ (or, effectively, two pions) that can be exchanged for the transitions $\jp\jp \to \jpsip$ or $\jpsipp$ to happen. In this regard, it is interesting to notice that indeed the dip prior to the $X(6900)$ peak appears around the $\jpsip$ threshold at 6783~MeV. Therefore, from this phenomenological point of view, among the double-charmonium channels, the $\jpsip$ and $\jpsipp$ ones are expected to play the most crucial role in describing the double-$\jp$ spectrum up to the energies covering the $X(6900)$ peak.

In this Letter, we aim at constructing minimal coupled-channel models able to describe the LHCb data on the double-$\jp$ invariant mass distribution in the energy interval from the double-$J/\psi$ threshold to 7.2 GeV and studying their predictions for pole locations and line shapes in the other double-charmonium channels. In particular, we consider a two- ($\jp\jp$ and $\jpsip$) and three-channel ($\jp\jp$, $\jpsip$, and $\jpsipp$) models and find that (i) both models provide a remarkably good description of the data which, therefore, do not allow one to distinguish between them, (ii) both models predict the existence of a near-threshold pole around 6.2~GeV (we call it the $X(6200)$) corresponding to a shallow bound or virtual $\jp\jp$ state, (iii) the structure of the other, above-threshold poles appears to be very different for the two models considered and so are the predicted line shapes in the $\jpsip$ channel. We conclude, therefore, that the existence of the $X(6200)$ is a robust consequence of the proposed coupled-channel approach, while additional measurements of the other double-charmonium channels are necessary in order to
better understand the nature of the higher poles.

{\it Coupled-channel model.}---Contrary to earlier attempts to understand the role played by the relevant double-charmonium thresholds for the double-$\jp$ spectrum~\cite{Wang:2020wrp}, 
the key idea of our approach is to present a minimal model consistent with the data and able to extract the poles responsible for the structures in the data. 
We, therefore, confine ourselves to those double-charmonium channels which are consistent with HQSS, and constrain the {$T$-matrix} with unitarity and causality.
Thus, we focus on two variants of the
coupled-channel model: a two-channel model employing $\{\jp\jp,\jpsip\}$
and a three-channel model using $\{\jp\jp,\jpsip,\jpsipp\}$. 

As detailed below we work with a separable potential $V$. Then
the $T$-matrix of the coupled-channel system can be written as
\begin{equation}
T(E) = V(E)\cdot[1-G(E)V(E)]^{-1},
\label{eq:T}
\end{equation}
where $E$ is the double-$\jp$ center-of-mass energy and $G$ is a diagonal matrix for the intermediate two-body propagators. We use the dimensionally regularized two-point scalar loop function~\cite{Veltman:1994wz}, 
\begin{align}
    G_i(E)=&\frac1{16\pi^2}\bigg\{a(\mu)+\log\frac{m_{i1}^2}{\mu^2}+\frac{m_{i2}^2-m_{i1}^2+s}{2s} \log\frac{m_{i2}^2}{m_{i1}^2} \nonumber\\
&+\frac{k}{E} \Big[ 
\log\left(2k_i E+s+\Delta_i\right) + 
\log\left(2k_i E+s-\Delta_i\right) \nonumber\\ 
& -  
\log\left(2k_i E-s+\Delta_i\right) - 
\log\left(2k_i E-s-\Delta_i\right)
\Big]\bigg\},\label{eq:GDR}
\end{align}
where $s=E^2$, $m_{i1}$ and $m_{i2}$ are the particle masses in the $i$-th channel, $\Delta_i= m_{i1}^2-m_{i2}^2$,  $k_i=\lambda^{1/2}(E^2,m_{i1}^2,m_{i2}^2)/(2E)$ is the corresponding three-momentum with $\lambda(x,y,z)=x^2+y^2+z^2 - 2xy - 2yz - 2xz$ for the K\"all\'en triangle function. Here $\mu$ denotes the dimensional regularization scale, and $a(\mu)$ is a subtraction constant. 
The $T$-matrix in Eq.~(\ref{eq:T}) respects the constraints of unitarity.

{Since the narrow dip and peak near 6.9~GeV are around the $\jpsip$ and $\jpsipp$ thresholds,
respectively, we consider only $S$ waves which are able to produce nontrivial threshold structures.} 
For the two-channel model, the potential $V$ is parameterized as
\begin{equation}
V_{\rm 2ch}(E)=\begin{pmatrix}
a_1 + b_1 k_1^2 & c \\
 c & a_2 + b_2 k_2^2
\end{pmatrix},
\label{eq:V2c}
\end{equation} 
where $a_{1,2}$, $b_{1,2}$, and $c$ are real free parameters. The momentum dependence of the potential is necessary here to produce nontrivial structures above the higher threshold, since purely constant contact-term potential can only produce bound or virtual state poles below threshold.

For the three-channel model, the potential $V$ is a $3\times 3$ matrix,
\begin{equation}
V_{\rm 3ch}(E)=\begin{pmatrix}
a_{11} & a_{12} & a_{13} \\
a_{12} & a_{22} & a_{23} \\
a_{13} & a_{23} & a_{33}
\end{pmatrix},
\label{eq:V3c}
\end{equation}
where $a_{ij}$'s are real parameters of the model, 
and, aiming at the simplest possible formulation of the model consistent with the data, we do not consider an explicit momentum dependence in this case. 

The production amplitude in the $\jp\jp$ channel (labelled as channel 1) can be constructed as
\begin{equation}
 {{\cal M}_1=P(E)\left[b + G_1(E)T_{11}(E) +\sum_{i=2,3} r_i G_i(E)T_{i1}(E)\right],}
\label{eq:M}
\end{equation}
where $T_{ij}$ are the elements of the $T$-matrix in Eq.~\eqref{eq:T}, {the ratios $r_i$ mimic potentially different production mechanisms for different channels ($r_3=0$ for the 2-channel fit), and the parameter $b\neq 1$ accounts for violation of unitarity in the
production mechanism which should be present in a 2-body treatment 
given the complexity of the inclusive reaction from which the data were
extracted.}
To describe the details of the short-distance production encoded in the function $P(E)$ above, we take it in an exponential form,
$P(E)=\alpha e^{-\beta E^2}$,
and fix the slope parameter $\beta=0.0123$~GeV$^{-2}$ from fitting to the double-parton scattering (DPS) distribution quoted in the LHCb paper~\cite{Aaij:2020fnh}. The energy dependence of the production operator accounts for the fact that the double-$\jp$ and $\jpsip$ two-particle systems can be produced at the parton level and interact before the final double-$\jp$ particles are detected. The overall strength parameter $\alpha$ is treated as a free parameter of the model.

Finally, the experimental double-$\jp$ distribution is fitted with the function $\rho(E)|{\cal M}_1|^2$, where $\rho(E)= k_1/(8\pi E)$ is the double-$\jp$ phase space factor. 

{\it Fit results.}---Before fitting the data we get rid of the parameters which weakly affect the distribution or can be recast into other constants. 
In particular, we set $\mu=1$~GeV and the subtraction constant in the loop function is fixed as $a(\mu=1~{\rm GeV})=-3$; its variance {within Eq.~\eqref{eq:T}} can be absorbed into the redefinition of the contact interactions in the potential. Also, we choose {the $r_i$-parameters in the amplitude (\ref{eq:M}) equal to 1 } since the fit does not call for their different values. 

\begin{figure}[tb]
\begin{center}
\includegraphics[width=\linewidth]{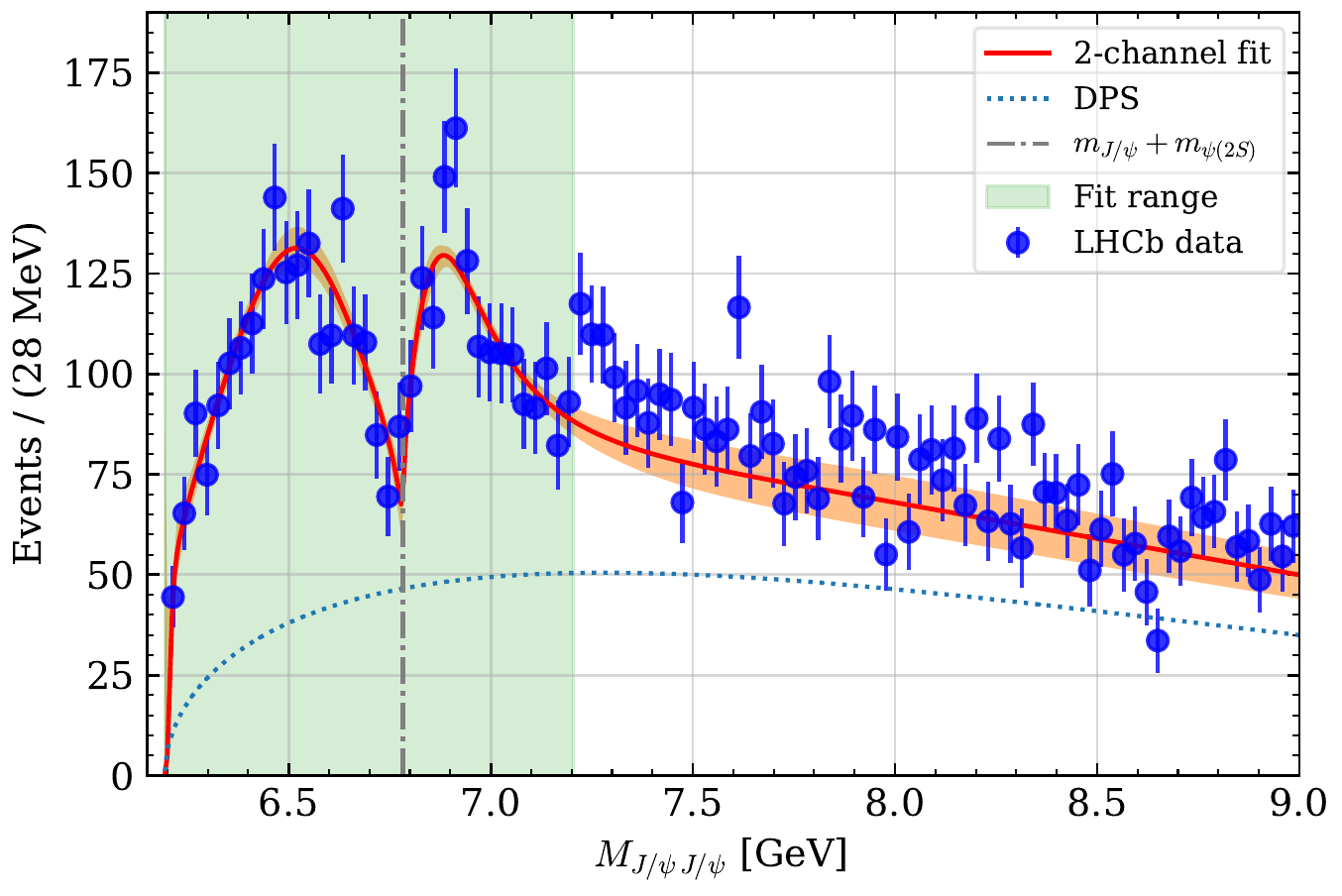}
\end{center}
\caption{Two-channel fit to the LHCb data of the double-$\jp$ invariant mass distribution~\cite{Aaij:2020fnh}. The solid line is the best fit with $\chisq=0.99$, and the band is the $1\sigma$ error area. The dotted line denotes $P(E)$ which perfectly describes the DPS distribution taken from the LHCb analysis~\cite{Aaij:2020fnh}.
} 
\label{fig:fit2c}
\end{figure}

{\bf Two-channel model:}
The two-channel parameterisation has 7 parameters. These are 
$\{a_1,a_2,b_1,b_2,c,{b},\alpha\}$. 
The fit was performed with randomly chosen $2\times10^4$ sets of initial values of the parameters and constrained by causality to ensure that there are no pole on the first Riemann sheet of the complex energy except on the real axis below threshold (see, e.g., Ref.~\cite{Gribov:2009zz}).
{The $\chi^2$ function is then minimised using the MINUIT algorithm~\cite{James:1975dr,iminuit,iminuit.jl}.} 
The best fit describes the data remarkably well with $\chisq=0.99$---see Fig.~\ref{fig:fit2c}.
Interestingly, although the fit was only performed up to 7.2~GeV, a good description of the data is achieved in the entire energy interval up to 9~GeV.
In this model, the dip in the line shape is produced due to a destructive interference of the $\jpsip$ threshold cusp which emerges from a coupled-channel dynamics with the background, see Eq.~\eqref{eq:M}. 
The above-threshold narrow hump is due to the energy dependence of the two-channel potential~\eqref{eq:V2c} which leads to a nearby resonance pole. A detailed analysis of the poles is given below.

\begin{figure}[tb]
\begin{center}
\includegraphics[width=\linewidth]{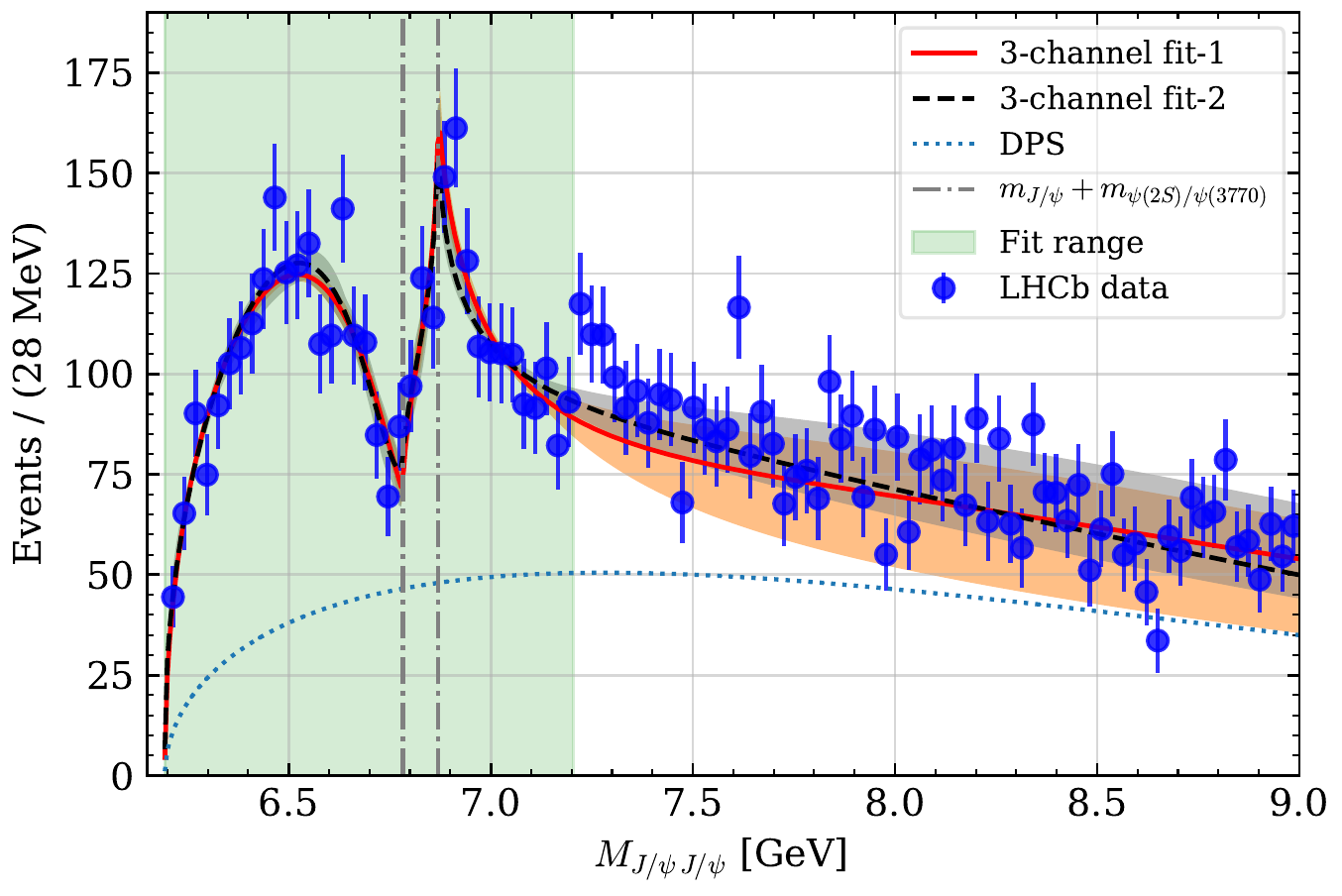}
\end{center}
\caption{Three-channel fits to the LHCb data~\cite{Aaij:2020fnh} of the double-$\jp$ invariant mass distribution. Fit~1 with $\chisq=0.97$ and Fit~2 with $\chisq=1.05$ are shown as the solid and dashed curves, respectively, together with the corresponding $1\sigma$ error bands. The dotted line is defined as in Fig.~\ref{fig:fit2c}. 
}
\label{fig:fit3c}
\end{figure}

{\bf Three-channel model:}
The three-channel model has 8 real parameters, $\{a_{ij}\,(i\geqslant j),{b},\alpha\}$. Two fits of similar quality are found with 
$\chisq=0.97$ (Fit~1) and $\chisq=1.05$ (Fit~2). All parameters of these fits coincide within their $1\sigma$ uncertainty except $a_{22}$. A comparison of both fits with the data is given in Fig.~\ref{fig:fit3c}. Like in the two-channel model, the description of the data is remarkably good, including the (not fitted) large-energy tail up to 9~GeV. In this model, the nontrivial structures in the line shape at approximately 6.8 and 6.9 GeV are due to the effect from the $\jpsip$ and $\jpsipp$ thresholds, amplified by a nearby pole.

\begin{figure}[tb]
\centering
\includegraphics[width=\linewidth]{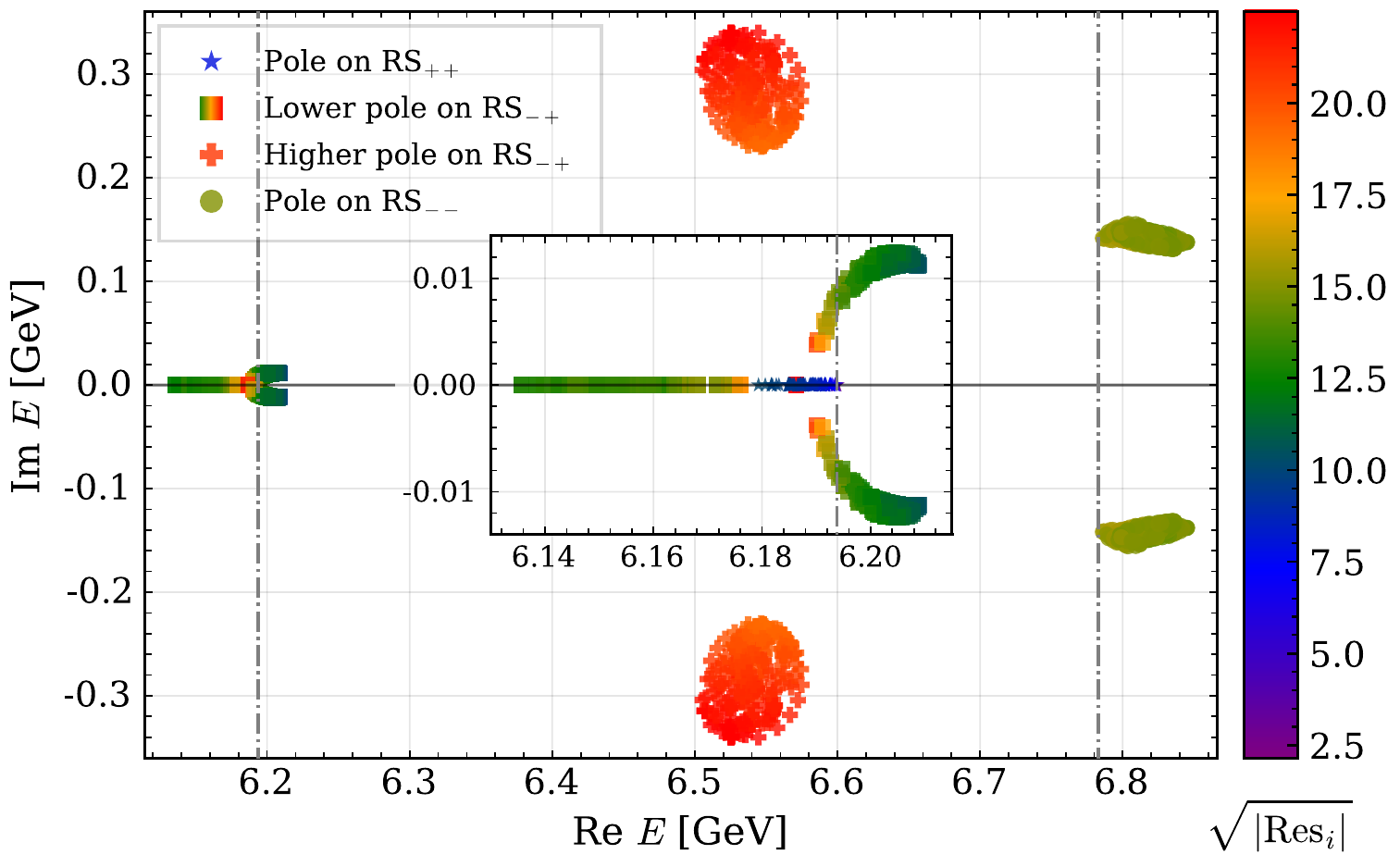}
\caption{Poles of the $T$-matrix from the 2-channel fit. The subplot zooms in the poles around 6.2~GeV. The effective couplings for the poles (namely, square-root of their residues) are encoded in color in the units of GeV. 
}
\label{fig:pole2c}
\end{figure}

\begin{figure}[tb]
\centering
\includegraphics[width=\linewidth]{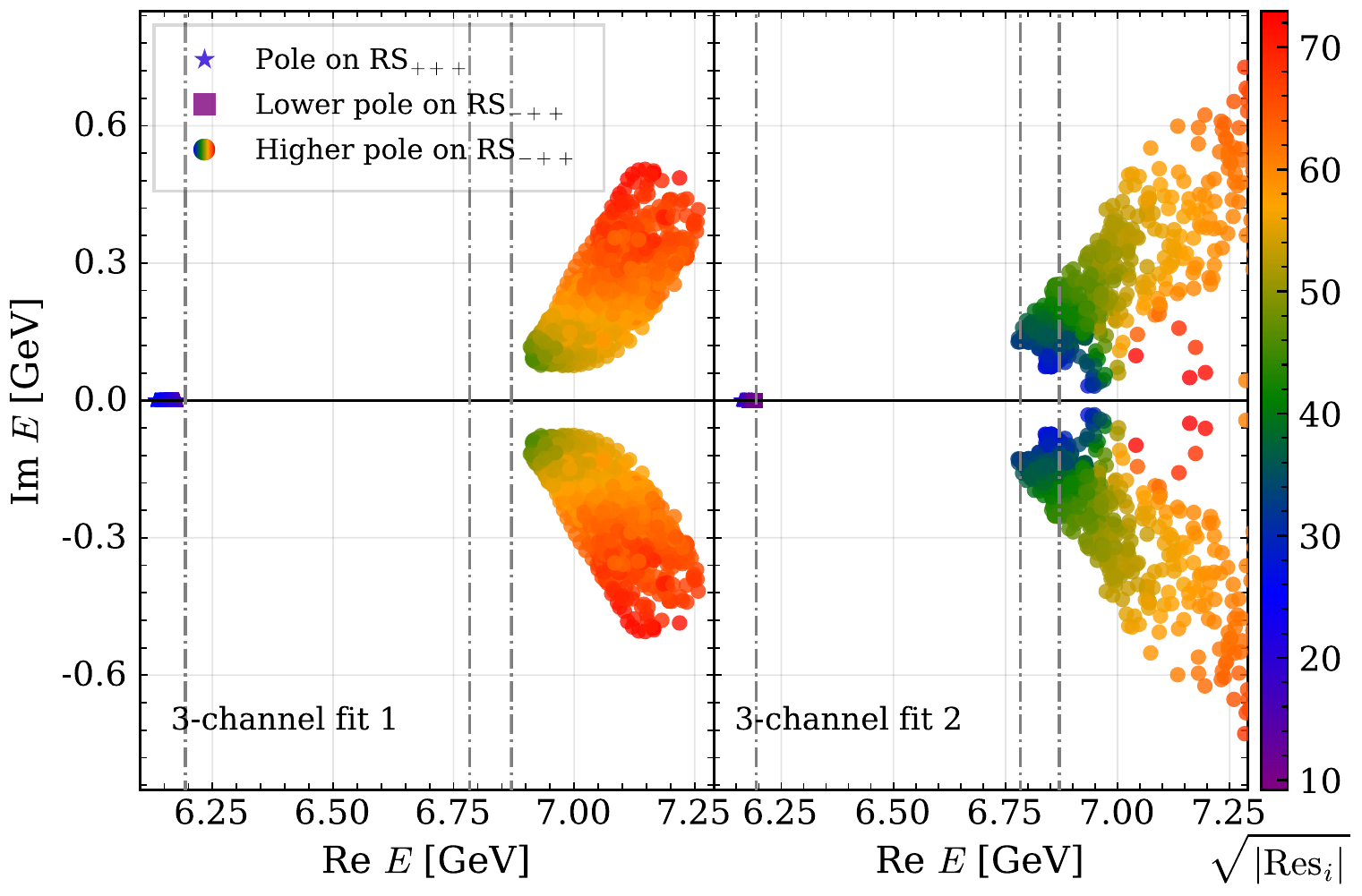}
\caption{Poles of the $T$-matrix from the 3-channel fits. {For the color coding, see the caption of Fig.~\ref{fig:pole2c}.}}
\label{fig:pole3c}
\end{figure}

{\it Pole analysis.}---To study the pole structure of the fitted $T$-matrix,
we generate more than 300 parameter sets within the $1\sigma$ contours in the parameter space for all combinations of the fit parameters and find all poles of the amplitude from the near-threshold region up to 7.2~GeV. The results are presented in Figs.~\ref{fig:pole2c} and \ref{fig:pole3c} from which one can draw several conclusions.

{We focus first on the mass region of the pronounced structures in the data
and study the $T$-matrix poles in the complex energy plane. For each pole we quote RS$_{\pm\pm\ldots}$ in parentheses to indicate the Riemann sheet where the pole is located, with the subscript composed of the signs of Im\,$k_i$ in all coupled channels involved, from the lowest in energy to the highest~\cite{Badalian:1981xj}.
We find that the pole locations are quite different for the different models employed. In particular, there exist two such poles for the two-channel model (hereinafter the pole positions are given in MeV),
\begin{eqnarray}
&&E_1^{\rm 2ch}=6542_{-36}^{+33}-i\, 282_{-52}^{+59}~ ({\rm RS}_{-+}),\nonumber\\[-2mm]
\\[-2mm]
&&E_2^{\rm 2ch}=6818_{-32}^{+28} - i\, 142_{-10}^{+14}~ ({\rm RS}_{--}),\nonumber
\end{eqnarray}
while, for the 3-channel fits, there is only one (badly determined) remote pole on RS$_{-++}$---see Fig.~\ref{fig:pole3c}.

Meanwhile, both models confidently predict a pole very near the $\jp\jp$ threshold. The two-channel fit allows for a virtual state, resonance, or bound state, 
\begin{equation}
  E_0^{\rm 2ch}=6203_{-27}^{+~6} - i\, 12_{-12}^{+~1}~ ({\rm RS}_{-+})~\text{or}~[6179, 6194]~({\rm RS}_{++}).
\end{equation}
The three-channel Fit 1 gives a bound state pole,
\begin{equation}
E_0^{\rm 3ch}[{\rm Fit~1}]=6163_{-32}^{+18}~ ({\rm RS}_{+++}),
\end{equation}
while Fit 2 allows either a shallow bound or virtual state,
\begin{align}
  E_0^{\rm 3ch}[{\rm Fit~2}]=6189_{-10}^{+~5}~ ({\rm RS}_{-++})~\text{or}~[6159, 6194]~({\rm RS}_{+++}). 
\end{align}
}

Obviously, further modifications of the model to extend the coupled-channel set or include higher-order terms in the potential cannot destroy this pole simply because such modifications would only affect the high-energy tail of the distribution, far away from the $\jp\jp$ near-threshold region. 
We conclude, therefore, that the existence of a pole near the $\jp\jp$ threshold is a robust consequence of the coupled-channel dynamics within the suggested approach. For definiteness, we name this state $X(6200)$. Its quantum numbers are either $0^{++}$ {($^{2S+1}\! L_J=^1\!\!S_0$)} or $2^{++}$ {($^5 S_2$)}, as required to have an $S$-wave threshold composed of two identical vector bosons. 
{As these two partial waves cannot interfere within our coupled-channel approach, we do not consider both amplitudes simultaneously. The latter scenario would require additional parameters not called for by the data.} 

\begin{figure}[tb]
\centering
\includegraphics[width=\linewidth]{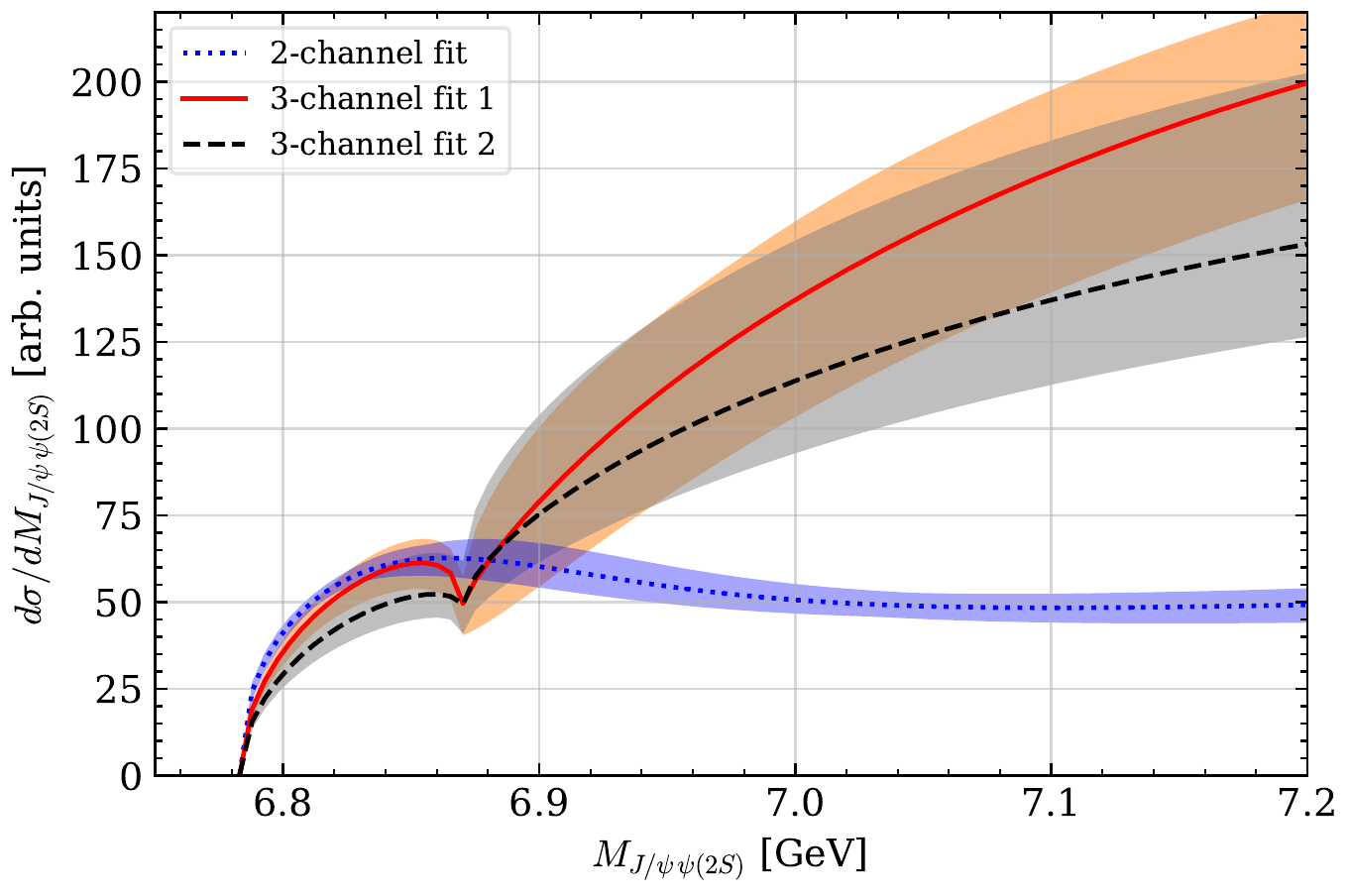}
\caption{Predictions for the invariant mass spectrum in the $\jpsip$ final state.}
\label{fig:prediction}
\end{figure}

{\it Further predictions and tests.}---As one can see from Figs.~\ref{fig:fit2c} and \ref{fig:fit3c}, although the models used to analyse the data in the $\jp\jp$ channel are based on a different dynamical content, they can provide a description of the data of a comparable quality. However, further predictions of these models differ substantially,
allowing for a direct experimental discrimination (or falsification of the whole approach).
As one of such tests we propose measurements of the line shapes in other double-charmonium channels. As a representative example, in Fig.~\ref{fig:prediction} we show the predictions of the two models employed in this work for the invariant mass spectrum in the $\jpsip$ final state. Indeed, the models predict quite different spectra above the $\jpsipp$ threshold, so that experimental data for this channel as well as for the $\jpsipp$ one should help to better understand the physical origin of the structures reported
by LHCb. 

Also, a direct experimental confirmation or refutation of the existence of the $X(6200)$ state is very important for a better understanding of the 
double-charmonium spectrum. In particular, a distinct signal from this state could be seen in the final states like $J/\psi\mu^+\mu^-$, $\mu^+\mu^+\mu^-\mu^-$, and  $\eta_c\eta_c$ which can be studied at energies below the nominal $\jp\jp$ threshold.

To better understand the nature of the $X(6200)$, {we estimate its compositeness,
$\bar X_A$ ($\bar X_A=1$ for molecules and $\bar X_A=0$ for compact states), which was introduced in Ref.~\cite{Matuschek:2020gqe} to characterize near-threshold bound states, virtual states, and resonances}. To this end, we employ the effective range expansion of the scattering amplitude in the $\jp\jp$ channel,
\begin{equation}
T(k) = - 8\pi \sqrt{s} \left[\frac1{a_0} + \frac12 r_0 k^2 -i\, k + \mathcal{O}(k^4) \right]^{-1},
\end{equation}
to extract the {$S$-wave scattering length $a_0$ and the effective range $r_0$}, and then use 
\begin{equation}
\bar X_A =(1+2|r_0/a_0|)^{-1/2}.
\end{equation}
The results presented in Table~\ref{Tab:ERE} imply
that while the two-channel model supports the $X(6200)$ as a compact state, the three-channel approach is better compatible with its molecular interpretation. 
If the latter is true, the same mechanisms {(for example, the two-pion exchange)} which drive the $X(6200)$ can provide sufficient binding also in other double-charmonium channels, so that many more double-charmonium molecular states can exist near relevant thresholds. 
In particular, if the $X(6200)$ is a molecule, HQSS predicts a near-threshold scalar double-$\eta_c$ state bound by a similar mechanism~\cite{Guo:2009id,Cleven:2015era}, to be searched for in, e.g., the $2[K\bar K]\pi$ final state.

\begingroup
\squeezetable
\begin{table}[t]
\caption{\label{Tab:ERE} The effective range parameters in the $\jp\jp$ channel and the compositeness $\bar{X}_A$ of the $X(6200)$. 
The sign of the scattering length by convention is negative (positive) if the $X(6200)$ is a bound (virtual) state.
}
\small
\begin{ruledtabular} 
\begin{tabular*}{0.48\textwidth}{@{\extracolsep{\fill}}cccc}
 & 2-ch. fit & 3-ch. fit 1 & 3-ch. fit 2 \\
 \hline
 $a_0 (\rm fm)$ & $\leq -0.49$ \!or\! $\geq 0.48$ & $- 0.61_{-0.32}^{+0.29}$ & $\leq -0.60$ \!or\! $ \geq 0.99$ \\
 $r_0 (\rm fm)$ & $-2.18_{-0.81}^{+0.66}$ & $-0.06_{-0.04}^{+0.03}$ & $-0.09_{-0.05}^{+0.08}$\\
 $\bar X_A$ & $\phantom{-}0.39_{-0.12}^{+0.58}$ & $\phantom{-}0.91_{-0.07}^{+0.04}$ & $\phantom{-}0.95_{-0.06}^{+0.04}$ \\
\end{tabular*}
\end{ruledtabular}
\end{table}

{\it Conclusions.}---In this Letter, we demonstrated that the recent LHCb data on the double-$J/\psi$ invariant mass spectrum are consistent with a coupled-channel description.
The best fits to the data imply the existence of a state near the $\jp\jp$ threshold which we called the $X(6200)$.
This state can have the quantum numbers of a scalar or a tensor. 
Further experimental tests are outlined to verify the hypothesis of the existence of this state and shed light on its nature. If confirmed, this discovery may start a new era in the spectroscopy of double-charmonium and double-bottomonium states. 
{It would also be valuable to simulate the double-$J/\psi$ (or double-$\eta_c$) scattering on the lattice.}

\bigskip

\begin{acknowledgments}

This work is supported in part by the National Natural Science Foundation of China (NSFC) under Grants No.~11835015, No.~11961141012 and No.~12047503, by the NSFC and the Deutsche Forschungsgemeinschaft (DFG, German Research Foundation) through the funds provided to the Sino-German Collaborative Research Center TRR110 ``Symmetries and the Emergence of Structure in QCD'' (NSFC Grant No. 12070131001, DFG Project-ID 196253076), by the Chinese Academy of Sciences (CAS) under Grants No.~XDB34030303 and No.~QYZDB-SSW-SYS013, and by the CAS Center for Excellence in Particle Physics (CCEPP). 
Work of V.B. and A.N. was supported by the Russian Science Foundation (Grant No. 18-12-00226).

\end{acknowledgments}

\begin{appendix}

    \section{Supplemental material}

Here we quote the values of the parameters found from the fits discussed in the main text and the corresponding correlation matrices. The parameters (without bars) used in the matrices $V_{\rm 2ch}$ and $V_{\rm 3ch}$ in the main text are obtained from the parameters (with bars) given here by multiplying the latter by the factor $\prod_{i=1}^4\sqrt{2m_i}$, where $m_i$'s are the involved charmonium masses for which we use \cite{Zyla:2020zbs}
\begin{equation}
m_{J/\psi}=3.0969~\mbox{GeV},\quad m_{\psi(2S)}=3.6861~\mbox{GeV}.
\end{equation}

In the convention used throughout the paper the $S$-wave $T$- and $S$-matrix are related as 
\begin{equation}
S_{ij}(E) = 1 - 2i \sqrt{\rho_i(E)\rho_j(E)} T_{ij}(E),\quad \rho_i(E) = \frac{k_i}{8\pi E},
\end{equation}
where $k_i$ is the absolute value of the c.m. 3-momentum in the $i$-th channel. 

The code employed is available on-line at \url{https://github.com/fkguo/double_jpsi_fit}.

\subsection{Two-channel model}

The fitted values of the parameters of the two-channel model are listed in Table~\ref{tab:2ch}. Their correlation matrix reads (see Table~\ref{tab:2ch} for the parameters order)
\begin{align}
 \begin{pmatrix}
 1 & 0.33 & 0.25 & -0.82 & 0.24 & -0.08 & 0.16 \\
 0.33 & 1 & -0.04 & -0.45 & 0.97 & -0.28 & 0.32\\
 0.25 & -0.04 & 1 & -0.50 & -0.21 & -0.19 & 0.25\\
 -0.82 & -0.45 & -0.50 & 1 & -0.35 & 0.45 & -0.55 \\
 0.24 & 0.97 & -0.21 & -0.35 & 1 & -0.30 & 0.33\\
 -0.08 & -0.28 & -0.19 & 0.45 & -0.30 & 1 & -0.99 \\
 0.16 & 0.32 & 0.25 & -0.55 & 0.33 & -0.99 & 1
 \end{pmatrix}.
\end{align}

The moduli of the $T$-matrix elements from the 2-channel fit are shown in Fig.~\ref{fig:tij2ch}.

\subsection{Three-channel model}

The fitted values of the parameters of the three-channel model are listed in Table~\ref{tab:3ch}. Their correlation matrices read (see Table~\ref{tab:3ch} for the parameters order)
\begin{align}
 \begin{pmatrix}
 1 & 0.94 & 0.57 & 0.82 & 0.56 & -0.38 & -0.20 & 0.07 \\
 0.94 & 1 & 0.74 & 0.93 & 0.70 & -0.52 & -0.47 & -0.23 \\
 0.57 & 0.74 & 1 & 0.92 & 0.98 & -0.93 & -0.50 & -0.37\\
 0.82 & 0.93 & 0.92 & 1 & 0.90 & -0.78 & -0.53 & -0.32 \\
 0.56 & 0.70 & 0.98 & 0.90 & 1 & -0.94 & -0.42 & -0.29 \\
 -0.38 & -0.52 & -0.93 & -0.78 & -0.94 & 1 & 0.41 & 0.34 \\
 -0.20 & -0.47 & -0.50 & -0.53 & -0.42 & 0.41 & 1 & 0.96\\
 0.07 & -0.23 & -0.37 & 0.32 & -0.29 & 0.34 & 0.96 & 1
 \end{pmatrix},
\end{align}
for Fit 1, and 
\begin{align}
 \begin{pmatrix}
 1 & 0.80 & 0.08 & 0.64 & 0.01 & 0.09 & 0.33 & 0.41 \\
 0.80 & 1 & 0.03 & 0.96 & -0.07 & 0.24 & -0.21 & -0.14 \\
 0.08 & 0.03 & 1 & 0.14 & 0.98 & -0.91 & 0.43 & 0.39\\
 0.64 & 0.96 & 0.14 & 1 & 0.04 & 0.15 & -0.33 & -0.29 \\
 0.01 & -0.07 & 0.98 & 0.04 & 1 & -0.92 & 0.46 & 0.41 \\
 0.09 & 0.24 & -0.91 & 0.15 & -0.92 & 1 & -0.48 & -0.42\\
 0.33 & -0.21 & 0.43 & -0.33 & 0.46 & -0.48 & 1 & 0.99\\
 0.41 & -0.14 & 0.39 & -0.29 & 0.41 & -0.42 & 0.99 & 1
 \end{pmatrix},
\end{align}
for Fit 2.

The moduli of the $T$-matrix elements from the 3-channel fits are shown in Fig.~\ref{fig:tij3ch}.

\begin{figure}[tbh]
 \centering
 \includegraphics[width=\linewidth]{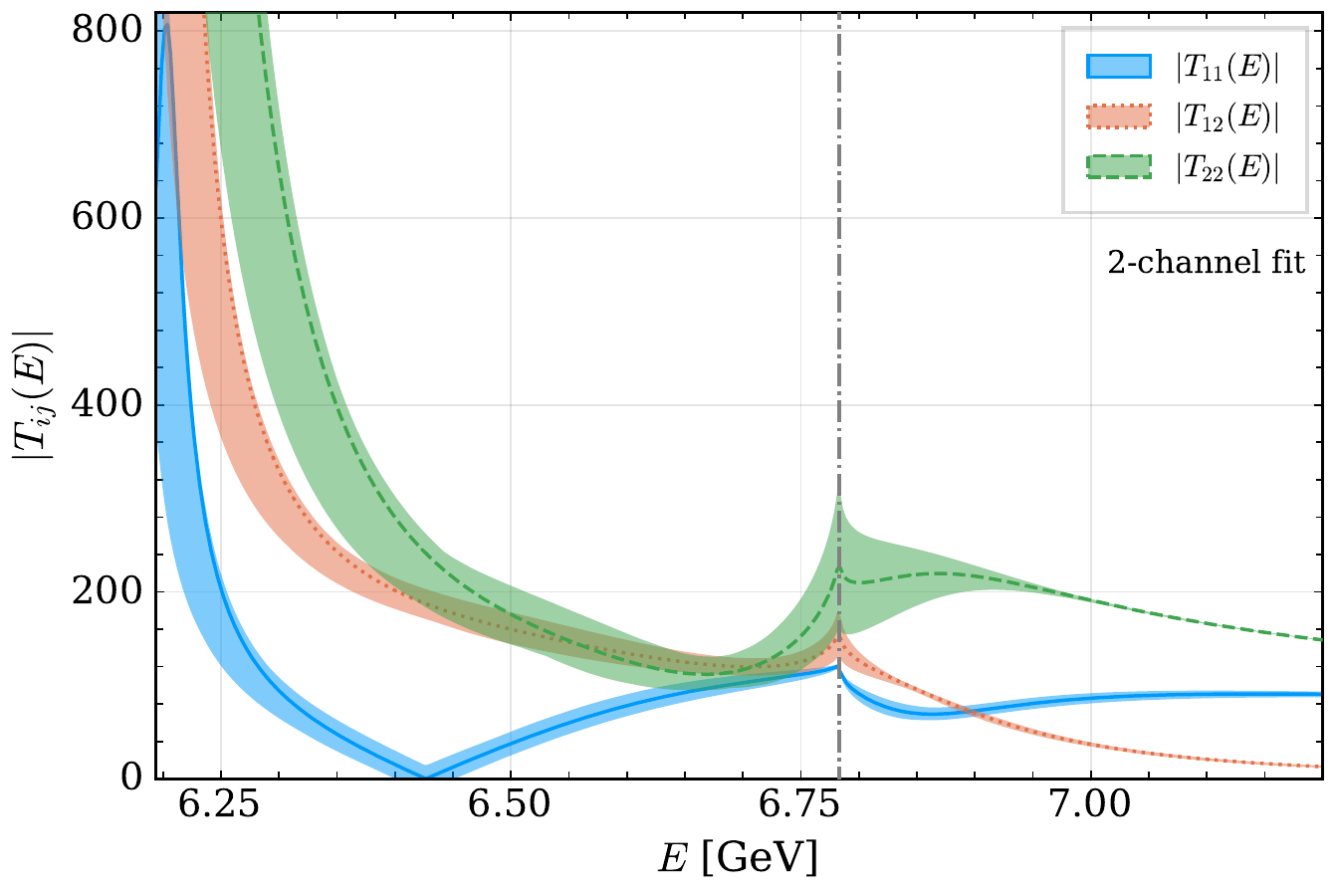}
 \caption{Moduli of the $T$-matrix elements from the 2-channel fit. The dash-dotted line show the $J/\psi\psi(2S)$ threshold.}
 \label{fig:tij2ch}
\end{figure}

\begin{figure}[tbh]
 \centering
 \includegraphics[width=\linewidth]{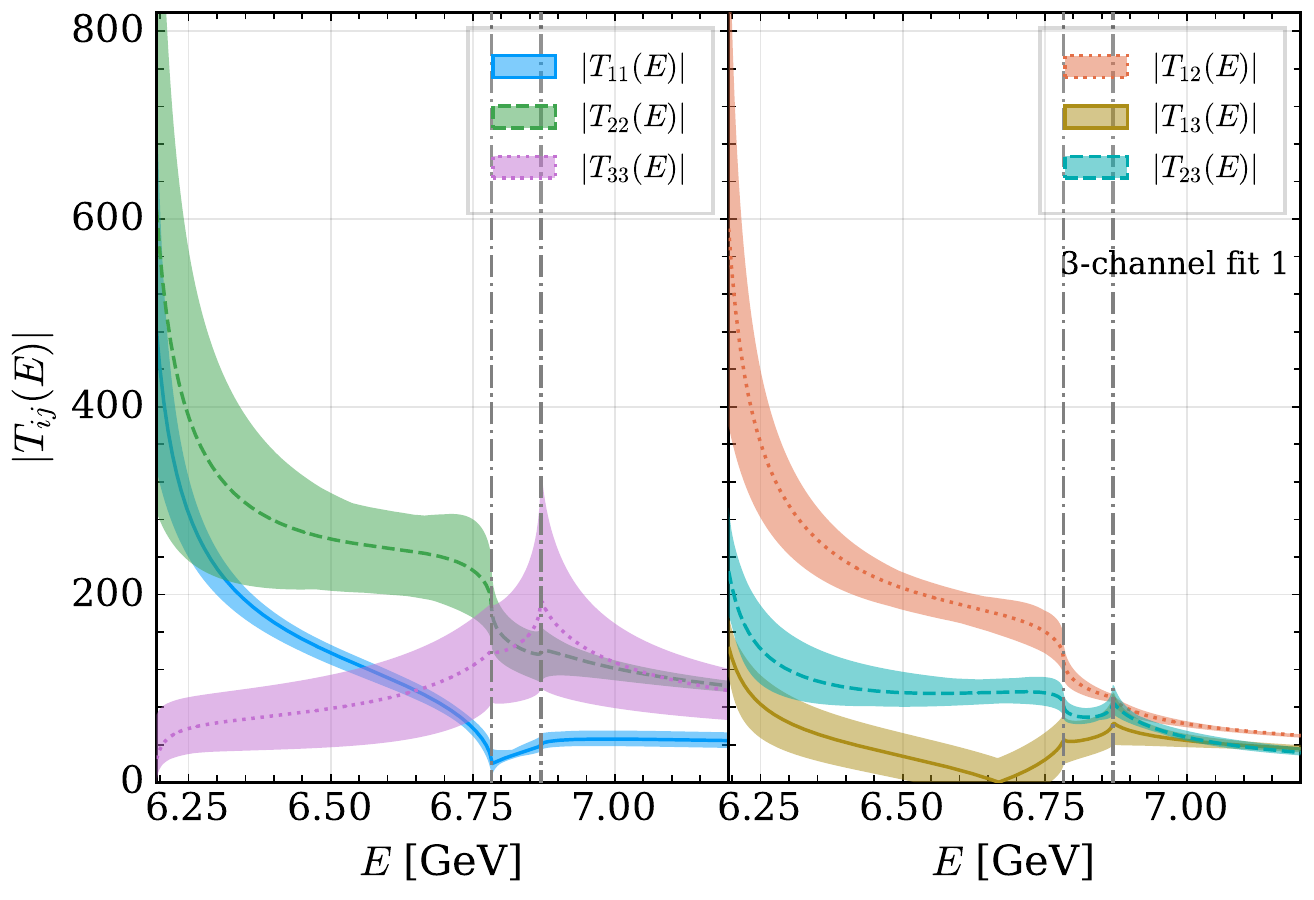}\\
 \includegraphics[width=\linewidth]{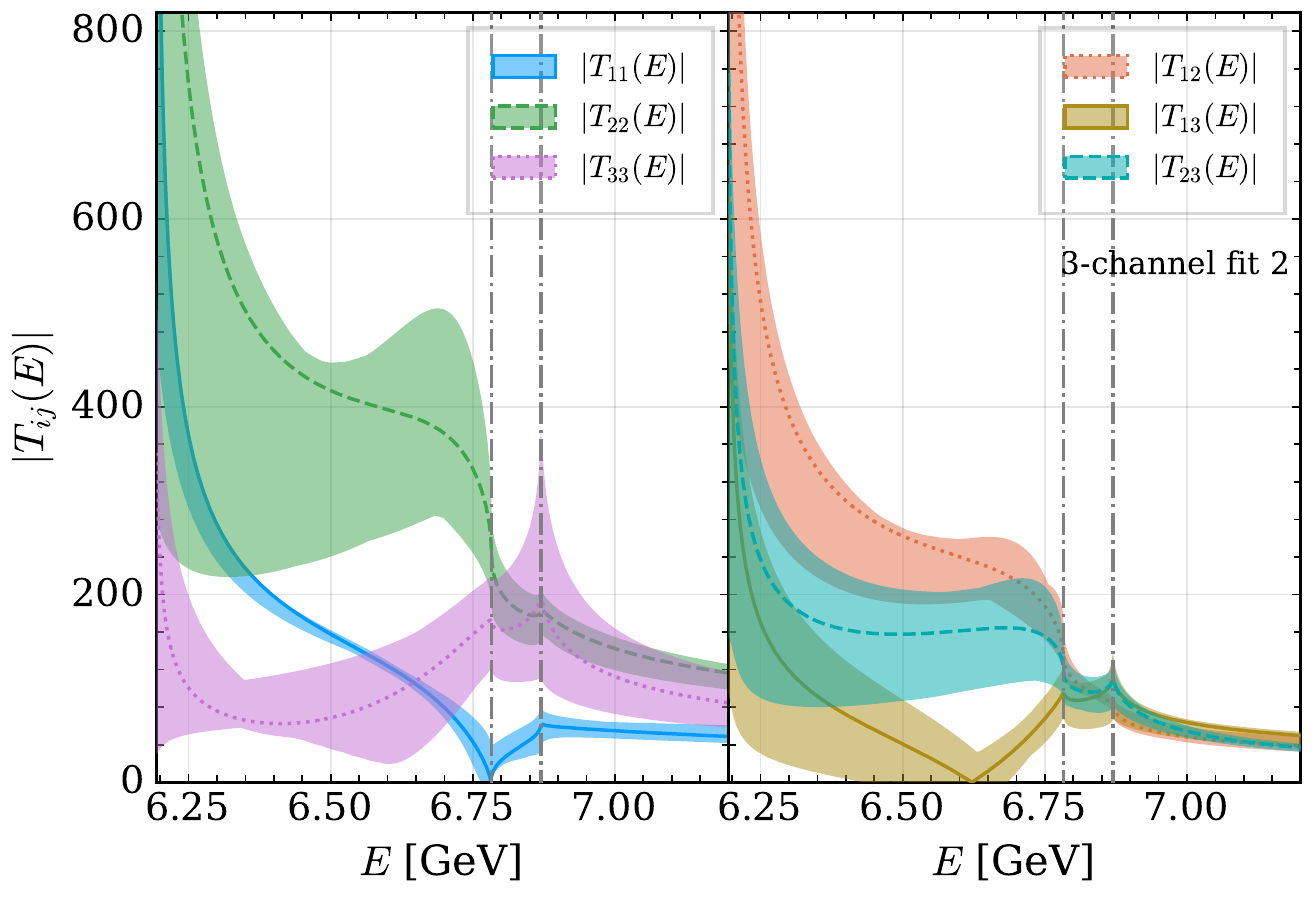}
 \caption{Moduli of the $T$-matrix elements from the 3-channel fits. The dash-dotted lines show the $J/\psi\psi(2S)$ and $J/\psi\psi(3770)$ thresholds.}
 \label{fig:tij3ch}
\end{figure}

\begin{table*}[t!]
    \caption{The values of the parameters for the two-channel model.}
    \begin{ruledtabular} 
    \begin{tabular}{ccccccccc}
    Parameters &$\bar a_1$ [GeV$^{-2}$] &$\bar a_2$ [GeV$^{-2}$]&$\bar c$ [GeV$^{-2}$]&$\bar b_1$ [GeV$^{-4}$]&$\bar b_2$ [GeV$^{-4}$]&$\alpha $&$b$\\
    \hline
    Fit&$0.2_{-0.5}^{+0.6}$&$-4.2\pm0.7$&$2.94_{-0.29}^{+0.36}$&$-1.8_{-0.5}^{+0.4}$&$-7.1\pm0.4$&$70_{-7}^{+8}$&$3.3\pm0.4$
    \end{tabular}
    \end{ruledtabular} 
    \label{tab:2ch}
    \caption{The values of the parameters for the three-channel model.}
    \begin{ruledtabular} 
    \begin{tabular}{cccccccccc}
    Parameters &$\bar a_{11}$ [GeV$^{-2}$]&$\bar a_{12}$ [GeV$^{-2}$]&$\bar a_{13}$ [GeV$^{-2}$]&$\bar a_{22}$ [GeV$^{-2}$]&$\bar a_{23}$ [GeV$^{-2}$]&$\bar a_{33}$ [GeV$^{-2}$]&$\alpha$&$b$\\
    \hline
    Fit 1 &$6.0_{-1.6}^{+2.2}$&$10.3_{-2.8}^{+3.4}$&$-0.2_{-1.3}^{+1.9}$&$13_{-4}^{+5}$&$-2.6_{-1.3}^{+2.4}$&$-2.3_{-1.1}^{+1.5}$&$ 
    250_{-60}^{+70}$&$-0.12_{-0.22}^{+0.21}$
    \\
    \hline
    Fit 2 &$7.8_{-2.0}^{+3.4}$&$16\pm4$&$0.9_{-2.5}^{+2.3}$&$26_{-~6}^{+12}$&$-3_{-5}^{+4}$&$-2.5_{-1.0}^{+2.1}$&$144_{-27}^{+67}$&$-0.7_{-0.4}^{+0.5}$\\
    \end{tabular}
    \end{ruledtabular} 
    \label{tab:3ch}
\end{table*}

\end{appendix}

\end{document}